\documentclass[reprint,superscriptaddress,amsmath,amssymb,aps,prb]{revtex4-2}
\usepackage{bm}
\usepackage{float}
\usepackage{graphicx}
\usepackage{mathtools}
\usepackage{braket}
\usepackage[usenames,dvipsnames]{color}
\usepackage[normalem]{ulem}
\usepackage[svgnames]{xcolor}
\usepackage{bm}
\usepackage{multirow}
\usepackage{titlesec}
\usepackage[utf8]{inputenc}
\usepackage{array}
\usepackage[colorlinks,linkcolor=blue,citecolor=blue,urlcolor=blue]{hyperref}
\usepackage{tabularx}
\usepackage{booktabs}
\usepackage{makecell}

\begin{document}
	
	\title{Thermal Hall resistivity and transverse entropy production in a phonon gas }
	
	\author{Xiaodong Guo}
	\affiliation{Wuhan National High Magnetic Field Center and School of Physics, Huazhong University of Science and Technology,  Wuhan  430074, China}

	\author{Xiaokang Li}
	\email{lixiaokang@hust.edu.cn}
	\affiliation{Wuhan National High Magnetic Field Center and School of Physics, Huazhong University of Science and Technology,  Wuhan  430074, China}

	\author{Alaska Subedi} 
	\affiliation{CPHT, CNRS, \'Ecole polytechnique, Institut Polytechnique
		de Paris, 91120 Palaiseau, France} 
	
	\author{Zengwei Zhu}
	\email{zengwei.zhu@hust.edu.cn}
	\affiliation{Wuhan National High Magnetic Field Center and School of Physics, Huazhong University of Science and Technology,  Wuhan  430074, China}
	
	\author{Kamran Behnia}
	\email{kamran.behnia@espci.fr}
	\affiliation{Laboratoire de Physique et d'\'Etude de Mat\'{e}riaux (CNRS)\\ ESPCI Paris, PSL Research University, 75005 Paris, France }
	
	\date{\today}
	
	\begin{abstract}
		Most theories of the phonon thermal Hall effect ignore phonon-phonon interactions. Here, by recalling the Senftleben-Beenakker effect in molecular gases, we argue that a magnetic field, by influencing collisions between neutral non-chiral [quasi-]particles, can induce a Hall response. Our study of two insulators with distinct crystal structures, layered honeycomb WS$_2$ and ferroelectric perovskite LiNbO$_3$, finds that  $\kappa_{xx}$ and $\kappa_{xy}$ peak at nearly the same temperature in both materials, as reported in other insulators. We show that the amplitude of transverse thermal \emph{resistivity} in clean and simple insulators is of the order of $|W_\perp/B|\simeq \frac{e}{k_B u}$, where $e$ and $k_B$ are fundamental constants and $u$ is the  binding energy density of the crystal. In complex and dirty insulators, $|W_\perp/B|$ is much larger and has a significant temperature dependence. Nevertheless, the peak thermal Hall \textit{angle} in all insulators remains roughly the same. 
		
	\end{abstract}
	\maketitle
	
	\section{Introduction}
	First observed two decades ago \cite{Strohm2005}, the phonon thermal Hall effect has since been reported in a broad variety of insulating solids \cite{Ideue2017,Sugii2017,Li2020,Grissonnanche2020,Boulanger2020,
		Akazawa2020,Sim2021,Chen2022,Uehara2022,Jiang2022,Li2023,Chen2024,
		Chen2024-2,Ataei2024,Meng2024,sharma2024phonon,Li2025,xiang2026}. Proposed microscopic descriptions fall into two broad categories: intrinsic mechanisms associated with phonon-band geometry and extrinsic mechanisms arising from skew scattering by defects \cite{Sheng2006,Zhang2010,Qin2012,Chen2020,Flebus2022,Guo2022} or by magnetic fluctuations \cite{Mangeolle2022}. Phonon--phonon collisions, which govern the longitudinal thermal conductivity \cite{Lindsay2013}, have rarely been invoked to explain the transverse response. 
	A possible reason is that electronic Hall effects can mostly be understood without invoking electron-electron interactions.  
	
	An instructive analogy is provided by the Senftleben–Beenakker (SB) effect in molecular gases \cite{Senftleben1930,Beenakker1962,kagan1967kinetic}. There, a magnetic field induces the precession of molecular angular momenta and renders opposite transverse momentum-transfer channels nonequivalent. Thus, intermolecular collisions generate an off-diagonal thermal conductivity even when the non-spherical molecules are not intrinsically chiral \cite{kagan1967kinetic,Mccourt1990}. The transverse response is therefore a consequence of field-modified particle--particle collisions rather than a single-particle Lorentz deflection. 
	
	We measured the longitudinal and transverse thermal conductivities, $\kappa_{xx}$ and $\kappa_{xy}$, in two nonmagnetic insulators with different crystal structures. The first, $2H$-WS$_2$, is a layered transition-metal dichalcogenide \cite{wilson1969transition}. The second, LiNbO$_3$, is a three-dimensional polar ferroelectric that hosts chiral phonons and provides a platform for developing quantum nondemolition (QND) measurements of phonons \cite{arrangoiz2018coupling,arrangoiz2019resolving,ueda2026chiral,ju2025lattice,Takuma2026}.
	In both materials, the temperature dependence of $\kappa_{xy}$ is closely linked to that of $\kappa_{xx}$, and the maximum thermal Hall angle approaches the upper scale identified across insulating crystals \cite{Li2023}. We further find that in both materials the empirical relation reported by Jin \textit{et al.} \cite{lishi2025} ($\kappa_{xy}\propto\kappa_{xx}^{2}$) approximately holds. By recasting the data in terms of the transverse thermal \textit{resistivity}, $W_{\perp} \equiv \nabla_yT/ J_q \simeq \kappa_{xy}/\kappa_{xx}^2$, we identify the flatness of the thermal Hall resistivity, $W_{\perp}$, as the underlying basis of this scaling. 
	
	We then demonstrate that this scaling is \textit{not} universal. It holds in simple clean insulators (silicon, germanium, black P and WS$_2$)  but not in complex and dirty insulators (such as cuprates or strontium titanate). Interestingly, LiNbO$_3$, which happens to be complex and clean, hosts a relatively large thermal Hall resistivity with no significant temperature dependence. We then show that the amplitude of a temperature-independent thermal Hall resistivity can be accounted for by a simple picture assuming that any change in the local energy density is accompanied by a concomitant change in the local charge density.
	
	Finally, comparing the thermal Hall response of a phonon gas with that of a molecular gas, we show that in an interacting picture, thanks to a transverse gradient in the frequency of entropy-producing events induced by the magnetic field, a thermal Hall response can arise even in the absence of phonon chirality \cite{juraschek2025chiral}.   
	
	\begin{figure*}[ht!]
		\centering
		\includegraphics[width=0.9\linewidth]{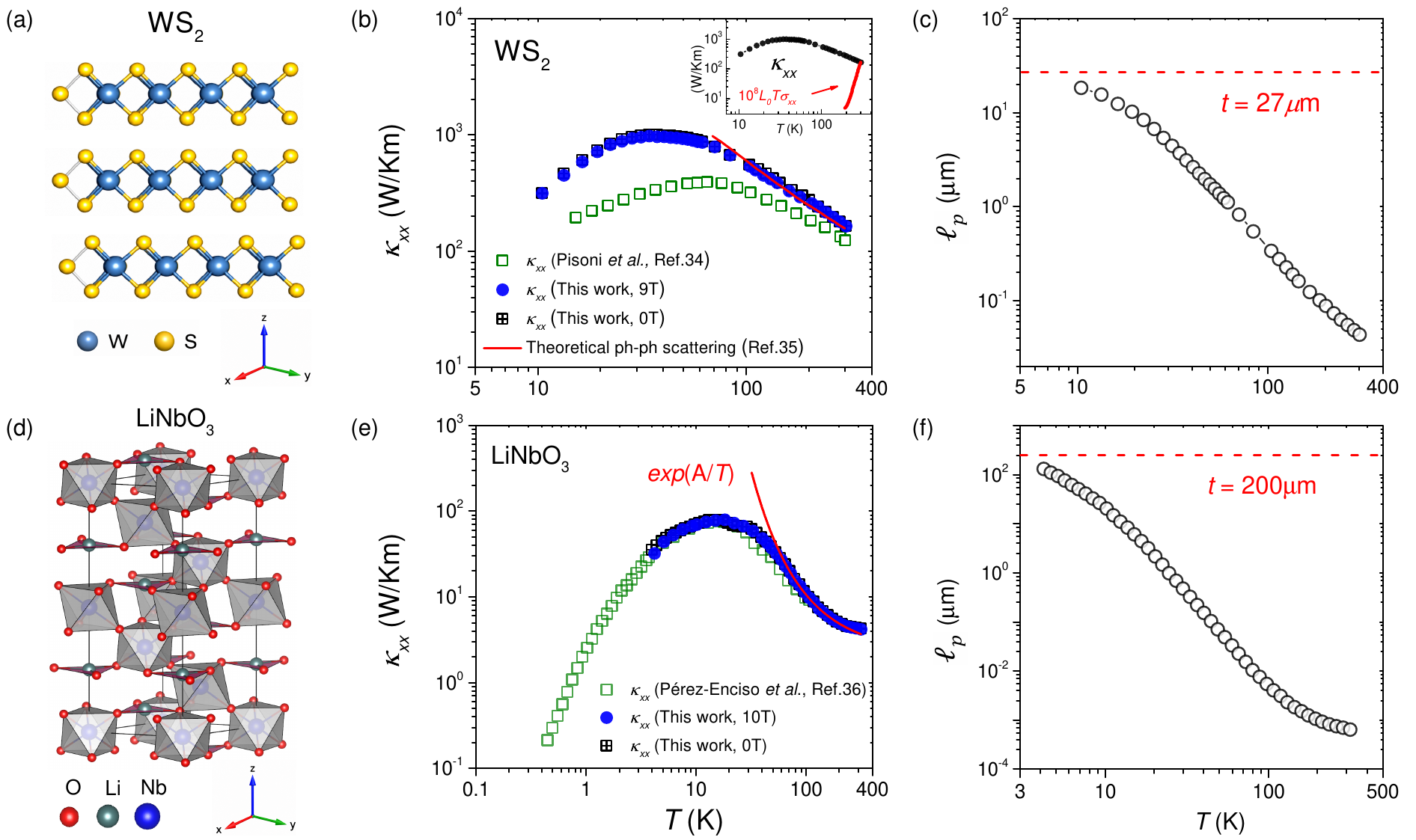} 
		\caption{\textbf{Crystal structures and longitudinal thermal transport of
				WS$_2$ and LiNbO$_3$.} (\textbf{a}) Crystal structure of layered $2H$-WS$_2$. (\textbf{b}) Temperature dependence of the longitudinal thermal conductivity $\kappa_{xx}$ of WS$_2$ at 0 and 9~T, together with previous results \cite{pisoni2016anisotropic} and theoretical calculations \cite{Lindroth2016}. The inset compares $\kappa_{xx}$ with $10^8L_0T\sigma_{xx}$, confirming the negligible electronic contribution to heat transport.
			(\textbf{c}) Phonon mean free path $\ell_p$ of WS$_2$, estimated from $\kappa_{xx}$, the specific heat, and sound velocity. The red dashed line marks the sample thickness, $t=27~\mu$m. (\textbf{d}) Crystal structure of LiNbO$_3$. (\textbf{e}) Temperature dependence of $\kappa_{xx}$ of LiNbO$_3$ at 0 and 10~T, compared with previous measurements \cite{PerezEnciso1998}. The red line represents an $\exp(A/T)$ dependence in the Ziman regime. (\textbf{f}) Phonon mean free path $\ell_p$ of LiNbO$_3$.}
		\label{fig:longitudinal}
	\end{figure*}
	
	\section{Results}
	
	\subsection{Longitudinal thermal conductivity} 
	We first establish the longitudinal phonon-transport characteristics of the two materials. $2H$-WS$_2$ is a layered diamagnetic semiconductor crystallizing in the hexagonal space group $P6_3/mmc$ [Fig.~\ref{fig:longitudinal}\textbf{(a)}] \cite{wilson1969transition,klein2001electronic,Lindroth2016}. Its electrical resistivity shows a semiconducting temperature dependence, and the electronic thermal conductivity, estimated from the Wiedemann--Franz law is approximately eight orders of magnitude smaller than the measured $\kappa_{xx}$ [inset of Fig.~\ref{fig:longitudinal}\textbf{(b)}]. Heat transport can therefore be regarded as overwhelmingly phononic throughout the temperature range of interest (Supplemental Material Note 1 \cite{SM}).
	
	The longitudinal and transverse temperature gradients were monitored simultaneously using three type-E thermocouples (Supplemental Material Notes 2 and 3 \cite{SM}). As shown in Fig.~\ref{fig:longitudinal}\textbf{(b)}, the longitudinal thermal conductivity of WS$_2$ is approximately $160$~W\,K$^{-1}$\,m$^{-1}$ at room temperature and reaches a maximum of nearly $1000$~W\,K$^{-1}$\,m$^{-1}$ at 34~K. The magnitude is substantially larger than that reported by Pisoni \textit{et al.} \cite{pisoni2016anisotropic} and closely approaches the first-principles prediction of Lindroth and Erhart \cite{Lindroth2016}. Our data, obtained in a presumably cleaner sample, suggest that disorder causes the discrepancy between previous data \cite{pisoni2016anisotropic} and the density functional theory (DFT) calculations \cite{Lindroth2016}.
	
	\begin{figure*}[ht!]
		\centering
		\includegraphics[width=0.95\linewidth]{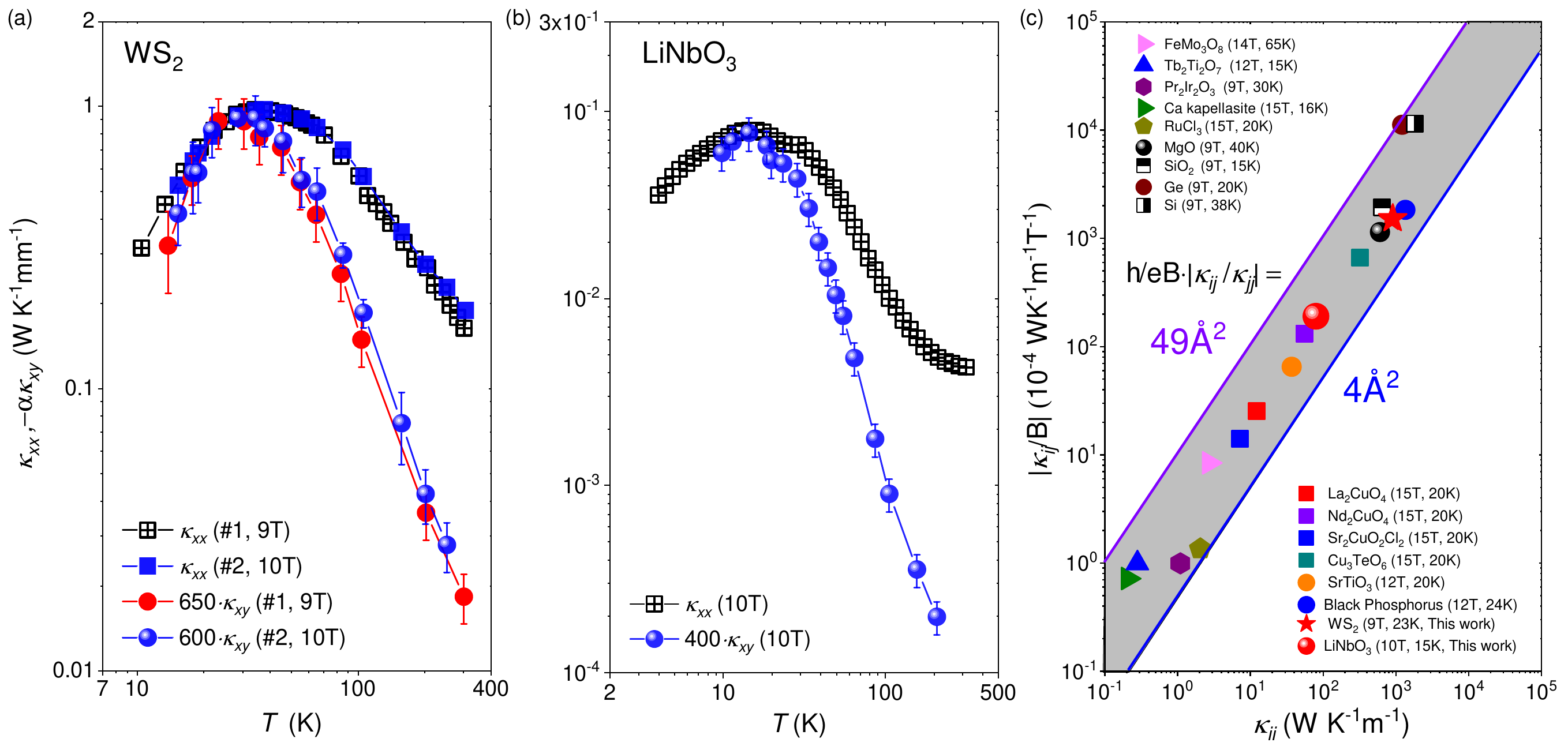}
		\caption{\textbf{Thermal Hall response in WS$_2$ and LiNbO$_3$.} (\textbf{a}) Temperature dependence of the longitudinal thermal conductivity $\kappa_{xx}$ of WS$_2$ sample~\#1, together with the transverse thermal conductivity $\kappa_{xy}$ of samples~\#1 and \#2 measured at 9 and 10~T, respectively. The $\kappa_{xy}$ data are multiplied by the factors indicated in the legend to emphasize their correspondence with $\kappa_{xx}$. (\textbf{b}) Temperature dependence of $\kappa_{xx}$ and $400 \cdot \kappa_{xy}$ of LiNbO$_3$ at 10~T. (\textbf{c}) Field-normalized transverse thermal conductivity $|\kappa_{ij}/B|$ plotted against the corresponding longitudinal thermal conductivity $\kappa_{ii}$ for different insulating crystals \cite{Ideue2017,Li2020,Grissonnanche2020,Boulanger2020,Akazawa2020,
				Chen2022,Uehara2022,Lefran2022,Li2023,Meng2024,lishi2025,Boulanger2022}.}
		\label{fig-transverse-data}
	\end{figure*}
	
	To quantify the phonon-scattering length scale, we estimate the phonon mean free path from $\ell_p\simeq\frac{3\kappa_{xx}}{C_vv_{\rm ph}}$, where $C_v$ is the lattice specific heat and $v_{\rm ph}$ is the in-plane acoustic velocity (Supplemental Material Note 4 \cite{SM}). From the acoustic slopes of the calculated phonon dispersion, we obtain $v_{\rm ph}\simeq5.7$~km\,s$^{-1}$ (Supplemental Material Note 5 \cite{SM}), consistent with the previously reported value \cite{Lindroth2016}. As shown in Fig.~\ref{fig:longitudinal}\textbf{(c)}, $\ell_p$ increases rapidly upon cooling and approaches the sample thickness, $t=27~\mu$m, near 10~K. The phonon transport therefore crosses into a boundary-limited, nearly ballistic regime at low temperatures.
	
	LiNbO$_3$ crystallizes in the polar rhombohedral space group $R3c$ below its Curie temperature  [Fig.~\ref{fig:longitudinal}\textbf{(d)}] \cite{ueda2026chiral}. Its thermal conductivity reaches a maximum of approximately $80$~W\,K$^{-1}$\,m$^{-1}$ near 20~K [Fig.~\ref{fig:longitudinal}\textbf{(e)}]. No resolvable difference is found between the 0 and 10~T data. On the high-temperature side of the maximum, $\kappa_{xx}$ follows an approximate $\exp(A/T)$ dependence over an extended interval, characteristic of Ziman heat transport \cite{kawabata2025}. The corresponding phonon mean free path, $\ell_p$, shown in Fig.~\ref{fig:longitudinal}\textbf{(f)}, increases rapidly upon cooling and approaches the sample thickness, $t=200~\mu$m, below  $\approx$ 3~K. LiNbO$_3$ therefore also enters a boundary-limited, nearly ballistic regime at the lowest temperatures. The observation of a finite thermal Hall signal in ferroelectric LiNbO$_3$ indicates that its absence in Ca-doped \cite{Jiang2022} and isotopically-enriched strontium titanate \cite{Sim2021} may arise from disorder, which is known to suppress the thermal Hall effect \cite{Xiang2026sto,ling2026phonon}, rather than from ferroelectric ordering.
	
	\subsection{Transverse thermal conductivity} 
	We now turn to the transverse thermal response. Figure~\ref{fig-transverse-data}\textbf{(a)} compares the longitudinal and transverse thermal conductivities of two distinct WS$_2$ crystals. In both samples, $\kappa_{xy}$  peaks near 27~K, modestly below the 34~K maximum of $\kappa_{xx}$. After multiplication by the constant factors indicated in the legend, the two $\kappa_{xy}(T)$ curves closely follow $\kappa_{xx}(T)$. A similar correspondence is observed in LiNbO$_3$ [Fig.~\ref{fig-transverse-data}\textbf{(b)}], where $\kappa_{xy}$ and $\kappa_{xx}$ reach their maxima within the same temperature interval. Such a proximity is similar to the available data in other insulators \cite{Li2023}. As seen in Fig. \ref{fig-transverse-data}\textbf{(c)}, the maximum amplitude of the thermal Hall angle in WS$_2$ and LiNbO$_3$ are consistent with the upper bound observed in other systems (see below). 
	
	\section{Discussion}
	
	\subsection {Amplitude and temperature dependence of the thermal Hall resistivity}
	
	The transverse thermal [Hall] resistivity is defined as $W_{\perp} \equiv \nabla_yT/ J_q$. In the low-field limit, $\kappa_{xy}\ll \kappa_{xx}$, one has $W_{\perp}\simeq \kappa_{xy}/\kappa_{xx}^2$.  Jin \textit{et al.} reported that in all insulators $\kappa_{xy} \propto \kappa_{xx}^2$. 
	
\begin{figure*}[ht!]
		\centering
		\includegraphics[width=1.0\linewidth]{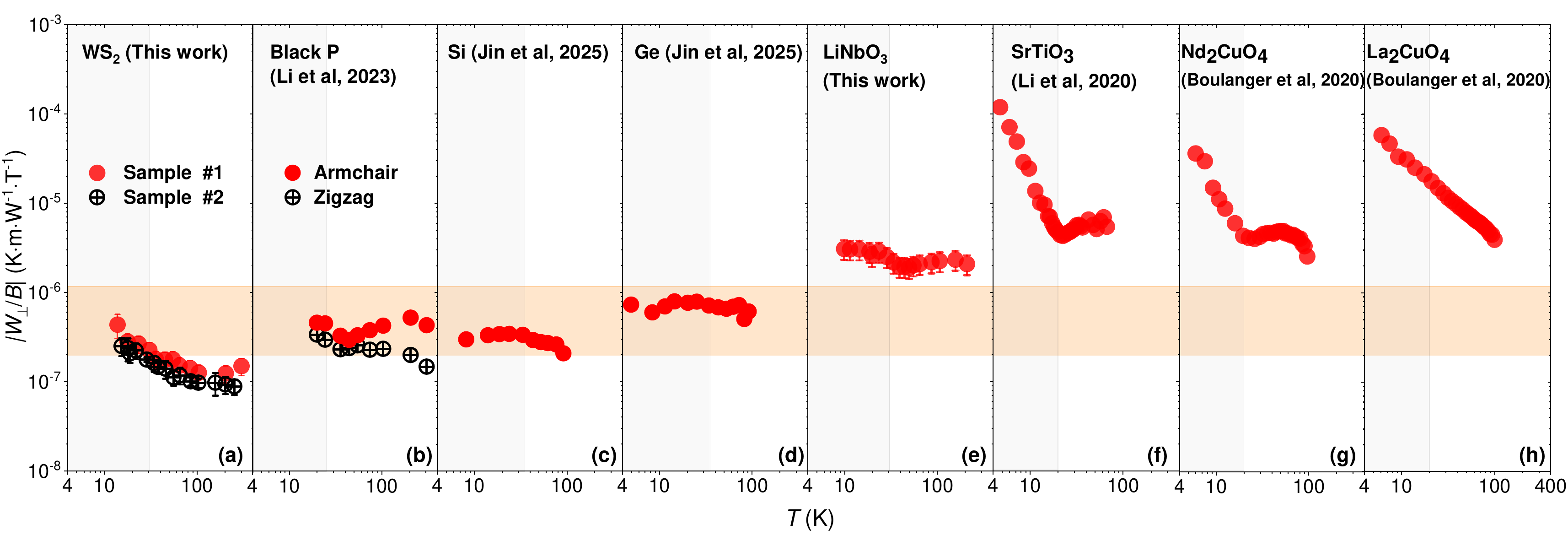} 
		\caption{\textbf{Field-normalized transverse thermal resistivity in different insulators.}
			Temperature dependence of $|W_{\perp}/B|=|(\nabla_y T/J_q^x)/B|$ in (a) two WS$_2$ crystals (this work); (b) black P with the heat current along the armchair and zigzag axes; (c) Si \cite{lishi2025}; (d) Ge \cite{lishi2025}; (e) LiNbO$_3$ (this work); (f) SrTiO$_3$ \cite{Li2020}; (g) Nd$_2$CuO$_4$ \cite{Boulanger2020}; and (h) La$_2$CuO$_4$ \cite{Boulanger2020}. The gray shaded regions indicate temperatures below the respective maxima of the longitudinal thermal conductivity $\kappa_{xx}$. The horizontal orange band represents the range predicted by Eq.~(3) for $\Gamma=1$ and $10^{10}\,\mathrm{Pa}<u<5\times10^{10}\,\mathrm{Pa}$.
		}
		\label{fig.resistivity}
	\end{figure*}

	Figure~\ref{fig.resistivity} compares the amplitude and the temperature dependence of the transverse thermal resistivity normalized by magnetic field $|W_{\perp}/B|$ in different insulators. The data on WS$_2$ (panel a) and LiNbO$_3$ (panel e) 
	obtained in this work are compared with those previously reported for other crystalline insulators \cite{Li2020,Boulanger2020,Li2023,lishi2025,ling2026phonon}. In clean insulators with ballistic phonons, $|W_{\perp}/B|$ exhibits a weak temperature dependence and its amplitude is confined to a relatively narrow window $\approx 10^{-7}-10^{-6}$~K$\cdot$m$\cdot$W$^{-1}$$\cdot$T$^{-1}$. In  complex oxides, such as SrTiO$_3$ \cite{Li2020}, Nd$_2$CuO$_4$, La$_2$CuO$_4$, and Sr$_2$CuO$_2$Cl$_2$ \cite{Boulanger2020}, neither of these two empirical facts holds. In these materials, $|W_{\perp}/B|$ is substantially larger than this lower bound. Moreover, instead of being almost flat as a function of temperature, it increases upon cooling.  
	
	\begin{figure}[t!]
		\centering
		\includegraphics[width=0.9\linewidth]{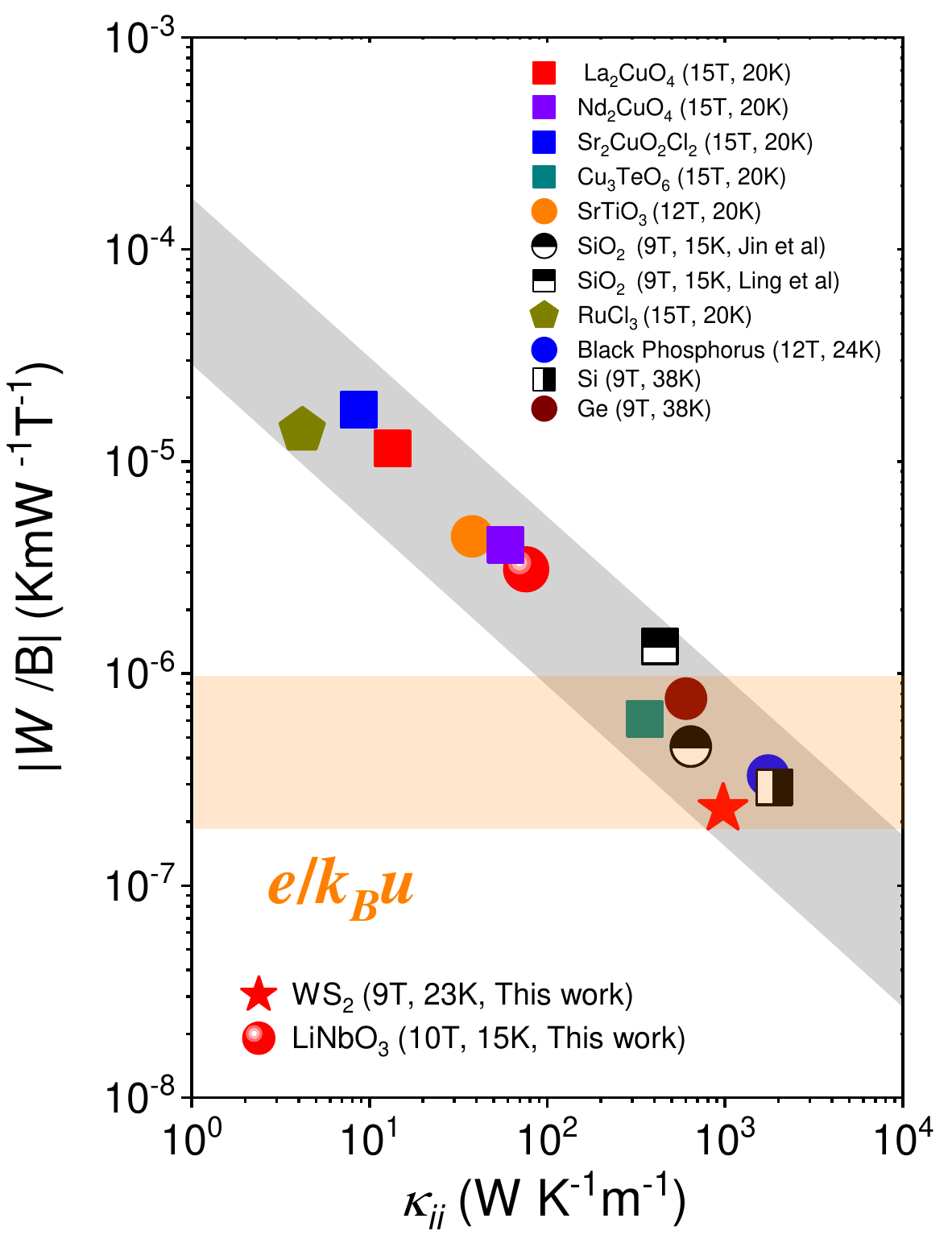}
		\caption{\textbf{Thermal Hall resistivity as a function of peak longitudinal thermal conductivity.} Field-normalized transverse thermal resistivity, $|W_{\perp}/B|$, plotted against the maximum $\kappa_{ii}$ in different crystalline insulators. Complex, dirty insulators (on the left) have a much larger $|W_{\perp}/B|$ \cite{Li2020,Boulanger2020,Akazawa2020,
				Chen2022,Lefran2022,Li2023,lishi2025,Boulanger2022}. }
		\label{fig-transverse}
	\end{figure}
	Thus, the scaling reported by Jin \textit{et al.} is not universal. Its validity is restricted to a subset of insulators, which are either elemental (silicon, germanium and phosphorus) or binary (WS$_2$ and SiO$_2$). This subset shares a common feature: their longitudinal thermal transport can be understood without invoking disorder. The peak in thermal conductivity separates a high-temperature phonon-phonon scattering regime and a low-temperature phonon-boundary scattering regime. Disorder does not play a significant role in either case. As we will see below, the amplitude of the thermal Hall resistivity of these insulators can be accounted for. In the second subset of insulators, on the other hand, disorder plays a significant role. Indeed, the phonon mean free path in complex oxides (SrTiO$_3$, Nd$_2$CuO$_4$, La$_2$CuO$_4$, and Sr$_2$CuO$_2$Cl$_2$) remains far below the sample size even at low temperatures.
	
	Fig. \ref{fig-transverse} displays the thermal Hall resistivity in different insulators as a function of the maximum longitudinal thermal conductivity. The change of focus from conductivity to resistivity clarifies one point: Dirtier systems (those with lower $\kappa_{ii}$, located on the left side of the figure) have a thermal Hall resistivity much larger than what is expected in a clean simple one. This may indicate the presence of an additional and distinct source of thermal Hall resistivity in them.
	
	Since the thermal Hall angle is the product of $|W_{\perp}/B|$ and $\kappa_{ii}$, the inverse correlation shown in Fig. \ref{fig-transverse} has another remarkable consequence. 
	
	\subsection{The maximum thermal Hall angle across insulators}
\begin{figure*}[ht!]
		\centering
		\includegraphics[width=1.0\linewidth]{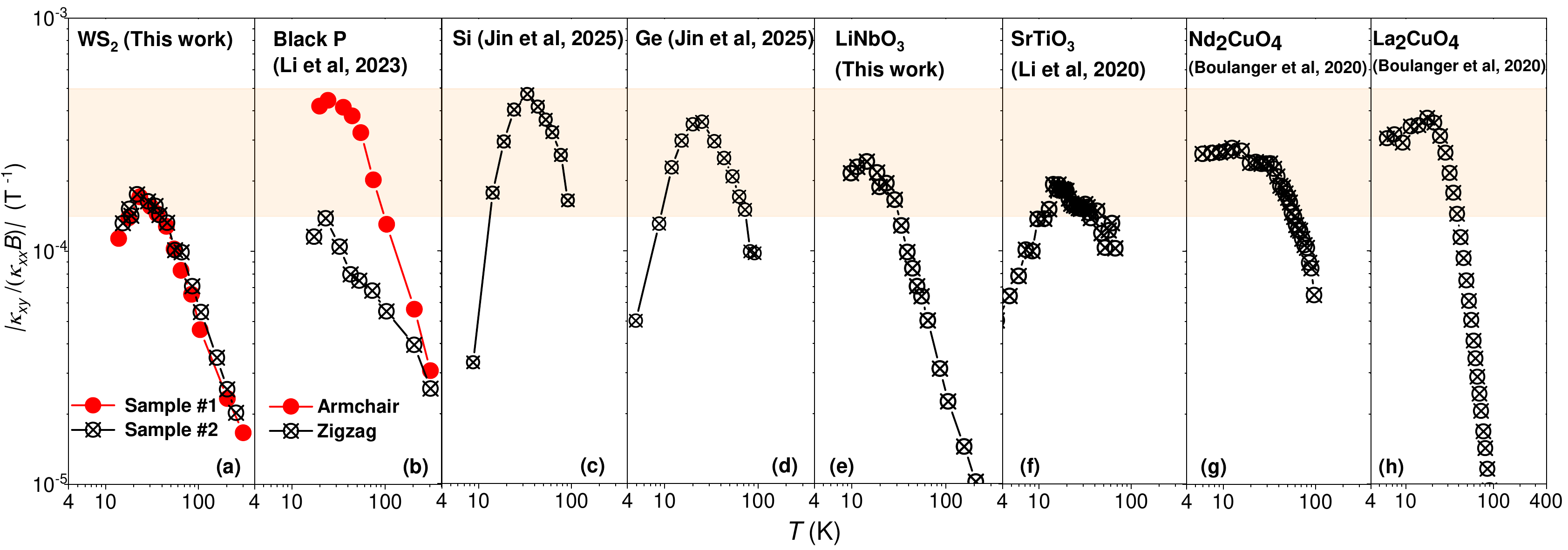} 
		\caption{\textbf{Transverse thermal Hall angle in different materials.}  (\textbf{a}) Thermal Hall angle $\frac{|\kappa_{xy}|}{\kappa_{xx}}$ normalized by magnetic field, $B$, as a function of temperature $T$ in WS$_2$.  Other panels show the same quantity for  black P \cite{Li2023} (\textbf{b}), Si \cite{lishi2025} (\textbf{c}), Ge\cite{lishi2025} (\textbf{d}), LiNbO$_3$ (\textbf{e}), SrTiO$_3$\cite{Li2020} (\textbf{f}), Nd$_{2}$CuO$_4$\cite{Boulanger2020} (\textbf{g}), and La$_{2}$CuO$_4$\cite{Boulanger2020} (\textbf{h}). }
		\label{Hall-angle}
	\end{figure*}

	Figure \ref{Hall-angle} shows the temperature dependence of the $|\kappa_{ij}/(B\kappa_{ii})|$ ratio in the same insulators shown in Fig. \ref{fig-transverse}. In all cases, this angle peaks at an intermediate temperature close to the peak temperature of longitudinal thermal conductivity. This is where hydrodynamic corrections to diffusive thermal transport are known to be strongest \cite{kawabata2025}, supporting the idea that Normal phonon-phonon scattering, which becomes relatively more frequent than Umklapp and boundary scattering near the peak, plays a significant role \cite{Behnia2025}.
	
	The most striking aspect of Fig. \ref{Hall-angle} is that, despite the wide variation in longitudinal and transverse thermal conductivities and resistivities, the peak thermal Hall angle remains within a remarkably narrow range: $|\kappa_{ij}/(B\kappa_{ii})|$ lies between $1.5\times 10^{-4}$ T$^{-1}$ and $5\times 10^{-4}$ T$^{-1}$. One of the authors of the present paper recently proposed \cite{Behnia2025} that this indicates the relevance of the field-induced geometric phase of phonons and two length scales in setting an upper bound to the thermal Hall angle. We do not discuss this feature further here. 
	
	\begin{table*}[htbp]
		\centering
		\caption{\textbf{Experimentally measured $|W_\perp/B|$ and Equation~\ref{eq:Wperp}.} The table lists insulating crystals together with their measured $|W_{\perp}/B|$, average velocity $\langle v_{s}\rangle$, mass density $\rho_M$ \cite{Green1990,Blakemore1982,Lange2007BlackP,Schutte1987WS2,Kihara1990Quartz,Weis1985LiNbO3}, and the estimated energy density $u=\frac{1}{2}\rho_M\langle v_{s}\rangle^2$. The quoted $|W_{\perp}/B|$ values are averaged over all data points. $\langle v_{s}\rangle$  is along the crystal axis of heat flow \cite{hall1967electronicSi,fujii1982inelastic-p,kushibiki2002accurate-Sio2,mcskimin1953measurement,Weis1985LiNbO3}, extracted from longitudinal, $v_l$, and transverse, $v_{ti}$ ($i=1,2$) sound velocities using: $3\langle v_{s}\rangle^{-3}=(v_l^{-3}+v_{t1}^{-3}+v_{t2}^{-3})$ \cite{ANDERSON1963909}. In the case of SiO$_2$, two different values were reported in two different studies \cite{lishi2025,ling2026phonon}.}
		\label{tab:comparison}
		\setlength{\tabcolsep}{10pt}
		\renewcommand{\arraystretch}{1.1}
		\begin{tabular}{lccccc}
			\toprule
			Material
			& $|W_{\perp}/B|$ & $\langle v_{s}\rangle$ & $\rho_M$ & $u= \frac{1}{2}\rho_M\langle v_s\rangle^2$ & $|\frac{W_{\perp}}{B}|/\frac{e}{k_B u}$\\
			& $\left(10^{-6}\, \mathrm{K\,m\,W^{-1}\,T^{-1}}\right)$ & $\left(\mathrm{km\,s^{-1}}\right)$
			& $\left(\mathrm{g\,cm^{-3}}\right)$ & $\left(10^{11}\,\mathrm{Pa}\right)$ & \\
			\midrule
			Si & 0.30  & 6.4 & 2.3 & 0.47 &  1.2  \\
			Ge & 0.69 & 3.9 & 5.3 & 0.40 & 2.4 \\
			Black P (zigzag) & 0.24 & 3.8 & 2.7 & 0.19 & 0.40  \\
			Black P (armchair) &  0.41 & 2.4 & 2.7 & 0.078 & 0.28  \\
			WS$_2$ & 0.21 & 2.5 & 7.6 & 0.24 & 0.44  \\
			SiO$_2$  & 0.50-1.3 & 5.0 & 2.7 & 0.34 & 1.5-3.8  \\
			LiNbO$_3$ & 2.4  & 4.2 & 4.6 & 0.41& 8.6  \\
			\bottomrule
		\end{tabular}
	\end{table*}
	
	\subsection {Origin of a temperature-independent $W_\perp/B$} 
	
	Independent of the microscopic details of the coupling between magnetic field and the phonon flow, the magnitude of the experimentally measured $W_{\perp}$ can be accounted for by a chain of simple arguments. 
	
	In equilibrium, a crystal is characterized by a distribution of total energy density $\epsilon(\mathbf r)$ and charge density $\rho(\mathbf r)$, both obtained by minimizing the same many-body energy functional. The total energy density is the sum of potential $u$ and kinetic energy $k$: $\epsilon = u+k$. Far below the melting temperature, $k\ll u$ and we can assume $\epsilon \simeq u$.
	Note that $k$ represents the energy density of the phonon gas and has a strong temperature dependence ($\propto T^4$ when $T \ll \Theta_D$). The much larger $u$ has almost no temperature dependence. 
	
	Ultimately, the binding energy of a crystal originates from Coulomb interactions. Phonons are charge neutral, but any imbalance in the phonon population leads to a local deformation of the lattice triggering a redistribution of electronic and nuclear charge. Therefore, a perturbation of local energy density induced by the applied external heat flux is expected to induce a modification of the local charge distribution. Let us assume a linear proportionality between them:
	\begin{equation}
		\delta q_{\rm eff}(\mathbf r,t)=\alpha\,\delta u(\mathbf r,t)
		\label{Eq-perturb}
	\end{equation}
	
	Here, $\alpha$ is a material-dependent electromechanical susceptibility and $\delta q_{\rm eff}$ is the effective charge density induced by $\delta u$. It does not require macroscopic charge transport.
	
	A finite heat flux, $J_q$, induces a phonon drift velocity of $J_q/k$, where $k$ is the energy density of phonons. Since phonons are neutral, however, this velocity does not directly couple to the magnetic field. On the other hand, there is another drift velocity, set by the ratio of the heat flux to the equilibrium binding-energy density:
	\begin{equation}
		v_d=\frac{J_q}{u}.
	\end{equation}
	
	This velocity characterizes the drift of $\delta q_{\rm eff}$, intimately linked to $\delta u$ through Eq. (1), and thereby couples the heat flux to a static magnetic field.  Under realistic experimental conditions (a heating power in the range of $\approx 0.01$ W applied to a crystal with $\approx$ 1 mm$^2$ cross section), the average elastic binding-energy density, $u$, being typically $10^{10}$--$10^{11}$ Pa \cite{deJong2015} yields a very slow drift velocity $v_d$ ($\approx 10^{-6}$ m/s). This is ten orders of magnitude smaller than the sound velocity and five orders of magnitude smaller than the drift velocity of phonons at 35 K, the peak temperature of thermal conductivity, in silicon. 
	
	Now, a magnetic field of $B$ exerts a transverse force on this extremely slow flow of electromechanical perturbation: $F^{B}_{\perp}= \tilde{a}e v_dB$. Here, $\tilde{a}=q_{\rm eff}/e$ is a dimensionless parameter and $e$ is the elementary charge. In steady state, this force is balanced by a transverse thermal force $F^{th}_{\perp}= \tilde{b} k_B\nabla_yT$, where $\tilde{b}$ is another dimensionless parameter.
	
	Equating the two transverse forces would lead us to the following simple expression for $W_{\perp}\equiv\frac{\nabla_yT}{J^x_q}$:
	
	\begin{equation}
		|W_{\perp}/ B|=
		\Gamma \frac{e }{uk_B}
		\label{eq:Wperp}
	\end{equation}
	
	The dimensionless parameter $\Gamma=\tilde{a}/\tilde{b}$ is a charge-to-entropy conversion parameter. 
	
	According to Eq. \ref{eq:Wperp}, $|W_{\perp}/B|$ is set by fundamental constants ($e/k_B= 1.16 \times 10^{4}$ K/V) divided by $u$, the internal binding energy density of the crystal and the material-dependent parameter $\Gamma$, which quantifies the amount of entropy in units of k$_B$ per fundamental charge $e$. Assuming the latter to be  $\sim \mathcal{O}(1)$, the typical value of $u\approx 10^{10}$--$10^{11}$ Pa, yields $|W_{\perp}/B|\approx  10^{-7}-10^{-6}$ K$\cdot$m$\cdot$W$^{-1}\cdot$T$^{-1}$, which is what the experiments found in simple clean insulators. 
	
	Note the similarity of Eq. \ref{eq:Wperp} to the established expression for thermal Hall resistivity of a generic electron gas: In both cases, $e/k_B$ is divided by an energy density. In the case of an insulating solid, $u$, the binding energy density,  has replaced  $\frac{\pi^2}{3} n k_BT$, the energy density of the electron fluid (see Supplementary Note 6 \cite{SM}).   
	
	Table \ref{tab:comparison} lists six solids together with their experimentally measured $|W_{\perp}/B|$, averaged over the measured temperature range, their mass density $\rho_M$, and their sound velocity $v_s$, from which a rough estimate of $u$ can be obtained. The expected amplitude of $|W_{\perp}/B|$ according to Equation \ref{eq:Wperp} is compared with the experimentally measured $|W_{\perp}/B|$. In silicon, germanium, black P, WS$_2$ and SiO$_2$, $|W_{\perp}/B|$ is of the same order of magnitude as $\frac{e}{uk_B}$, which implies   $\Gamma \simeq1$. Only in complex [ferroelectric and piezoelectric] LiNbO$_3$,  $\Gamma$ is significantly larger than unity. In all these solids, disorder is irrelevant to longitudinal thermal transport.
	
	In contrast to this subset of insulators, in complex oxides, such as cuprates, $\alpha-$RuCl$_3$ or strontium titanate, not only $|W_{\perp}/B|$ is more than an order of magnitude larger than $\frac{e}{uk_B}$, but also it presents a substantial temperature dependence (See Fig. \ref{fig-transverse-data}). In these solids, which host phonons with a short mean free path, the thermal Hall resistivity can only be accounted for by invoking either a large and temperature-dependent $\Gamma$ or the presence of another, possibly extrinsic, source of the thermal Hall signal.
	
	\subsection {The phonon gas compared to the molecular gas} 
\begin{figure*}[ht!]
		\centering
		\includegraphics[width=0.95\linewidth]{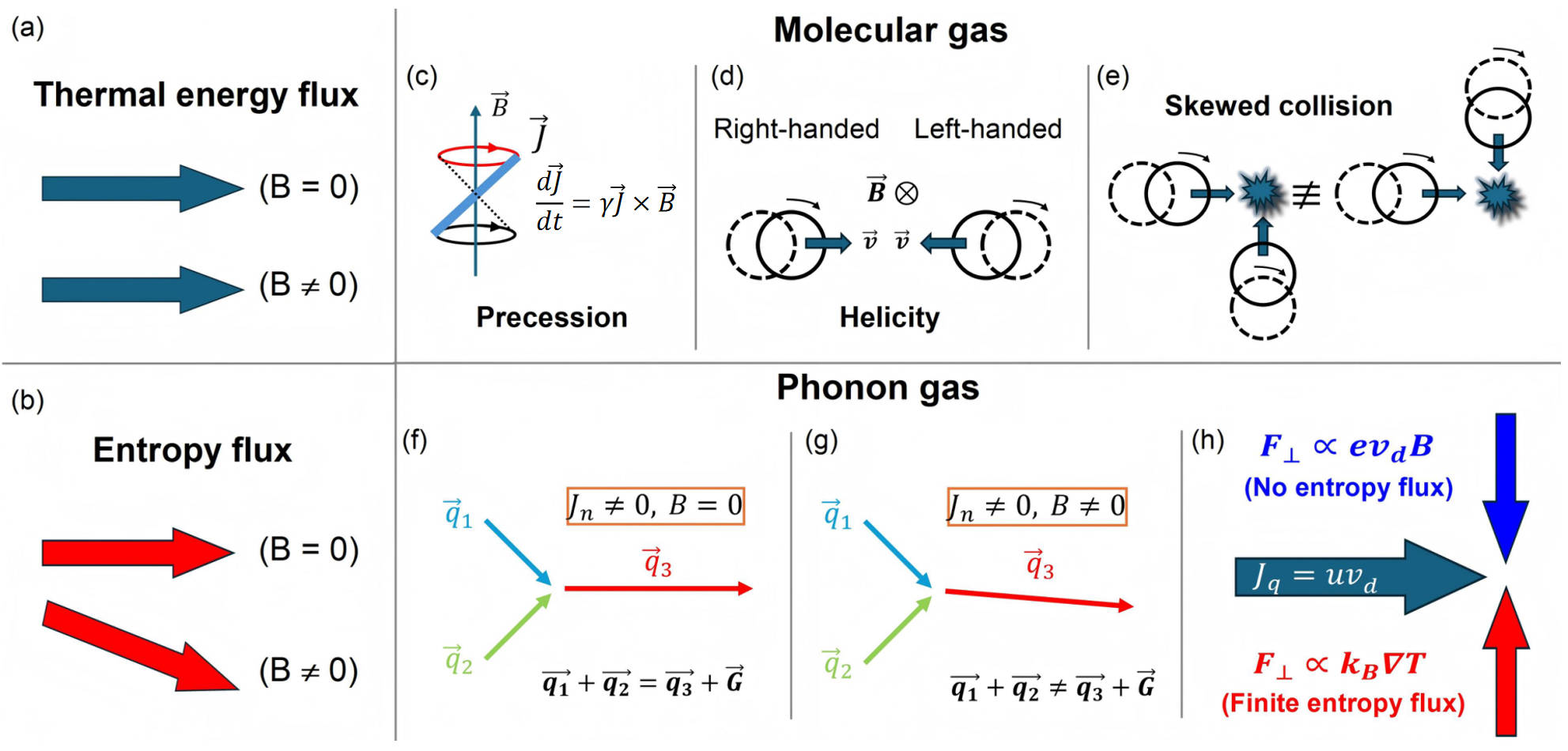} 
		\caption{\textbf{Field-induced twist of entropy flow in gases consisting of neutral [quasi-]particles.} A magnetic field does not deflect the thermal-energy flux (\textbf{a}), but twists the flow of entropy (\textbf{b}) by generating a transverse gradient in the density of entropy-producing events in a molecular gas (\textbf{c-e}) as well as in a phonon gas (\textbf{f-h}). (\textbf{c}) A magnetic field induces a precession of the molecular angular momentum $\vec{J}$, governed by $d\vec{J}/dt=\gamma\vec{J}\times\vec{B}$. (\textbf{d}) This precession combined with translational motion produces right- and left-handed dynamics. (\textbf{e}) In the presence of a magnetic field, the total angular momenta of colliding molecules lead to a difference in collision cross section and a skewed transverse momentum transfer. (\textbf{f}) In a phonon gas, the particle number is not conserved and phonons are created and destroyed during three-phonon events, which conserve pseudomomentum, modulo $\textbf{G}$, a unit vector of the reciprocal lattice. (\textbf{g}) In the presence of a magnetic field, this conservation law is altered due to a change in the electromagnetic pseudomomentum before and after scattering events. (\textbf{h}) Ultimately, this arises because the two transverse forces opposing each other do not have the same entropy content.}
		\label{fig.two gas}
	\end{figure*}

The phonon gas, like the molecular gas and unlike the electron gas, is a gas of neutral [quasi-]particles. Its thermal Hall response is \textit{not} a consequence of the bent trajectory of charged particles between collisions. It arises because a magnetic field, which does no work [Fig. \ref{fig.two gas}\textbf{(a)}], alters the details of entropy-producing events and generates a flow of entropy perpendicular to the applied heat flow [Fig. \ref{fig.two gas}\textbf{(b)}].
	
	The transverse thermal conductivity $\kappa_{xy}$ in a gas of non-spherical molecules is given by \cite{hermans1970transverse}: 
	\begin{equation}
		\frac{\kappa_{xy}}{\kappa_0} ({\rm{gas}}) \propto \frac{ \Theta}{1 + \Theta^2}.
		\label{EQ2}
	\end{equation}
	
	Here $\Theta = \omega_L \tau$, $\omega_L$ is the precession frequency of the molecular rotational angular momentum, and $\tau$ is the molecular scattering time \cite{HERMANS196781}. Equation \ref{EQ2} implies that the transverse thermal conductivity is maximal for $\Theta \simeq 1 $, vanishes when $\Theta \gg 1$, and when $\Theta \ll 1$ becomes linear in $\Theta$. Note that molecules are not required to be `chiral', just non-spherical.
	
	Collisions between molecules play a key role in the emergence of a transverse thermal gradient in a real gas. A magnetic field, by inducing Larmor precession governed by $d\vec{J}/dt=\gamma\vec{J}\times\vec{B}$ [Fig. \ref{fig.two gas}\textbf{(c)}], reduces the angular distribution of the angular momenta. The sense of precession combined with the orientation of linear momentum yields handedness to the molecules [Fig. \ref{fig.two gas}\textbf{(d)}]. Now, when a molecule collides with another molecule, the collision cross section depends on the orientation of the velocity of the incoming molecule with respect to the magnetic field. The pair of colliding molecules has a combined angular momentum $\vec{L}=\mu\vec{r} \times\vec{v}$, where $\mu$ is the reduced mass of the two molecules. Since angular momentum is conserved, the orientation of $\vec{L}$ with respect to the magnetic field leads to a difference in collision cross sections between the two lateral orientations for an incoming molecule [Fig. \ref{fig.two gas}\textbf{(e)}]. 
	
	Thus, the magnetic field and the heat flow conspire to generate a Hall response out of non-chiral molecules. The main parameters are the frequency of the Larmor precession, $\omega_L$, and the collision time, $\tau$. The process is most efficient when $\omega_L \tau\simeq 1$. Experimentally, the thermal Hall conductivity of molecular gases displays a universal non-monotonic dependence on the ratio of magnetic field to pressure, $\frac{H}{p}$ \cite{HERMANS196781}. When pressure, which sets $\tau$, and magnetic field, which sets $\omega_L$, optimize the interplay between thermal homogenization and precession of angular momenta, the signal is at its peak.
	
	Maksimov and Khabarova \cite{Maksimov2008} were the first to draw an analogy between the Senftleben–Beenakker effect in molecular gases and the phonon thermal Hall effect in solids. 
	
	In a phonon gas, the particle number is not conserved. Hence, a temperature gradient generates a net phonon density flux $\vec{J}_n$ \cite{hook2013solid}, in contrast to a real gas. Three-phonon processes are emissions and absorptions during which the phonon number decreases or increases. In the absence of a magnetic field, discrete spatial translational symmetry yields a pseudomomentum balance condition for any three-phonon interaction: $\mathbf{q}_1 + \mathbf{q}_2 = \mathbf{q}_3 + \mathbf{G}$ where $\mathbf{q}_i$ are the wavevectors of the participating phonons and $\mathbf{G}$ is a reciprocal lattice vector [Fig. \ref{fig.two gas}\textbf{(f)}]. In the presence of a magnetic field, the non-vanishing field-matter pseudomomentum density (the Minkowski pseudomomentum), $\mathbf{B} \times \mathbf{E}$ \cite{Nelson1991,Thellung1994,Streib2021}, enters the global momentum conservation tensor. Since the lattice displacement field before and after scattering events is not necessarily identical, the microscopic balance condition is altered, and the outgoing phonon wavevector $\mathbf{q}_3$ deviates from the linear sum of the incoming wavevectors $\mathbf{q}_1 + \mathbf{q}_2$ [Fig. \ref{fig.two gas}\textbf{(g)}].
	
	A formal and rigorous solution of the problem requires solving the Boltzmann transport equation with the collision integral terms on the right-hand side modified by the finite magnetic field (see Supplementary Note 7). However, even in the absence of this treatment left for future theoretical investigations, one can see that the flow of perturbative electric charge induced by heat flux will generate two transverse forces countering each other [Fig. \ref{fig.two gas}\textbf{(h)}]. Since the entropy content of these two forces is unequal, the entropy flow is twisted in a magnetic field [Fig. \ref{fig.two gas}\textbf{(b)}]. 
	
	\section{Concluding remarks}
	In summary, we found that in both WS$_2$ and LiNbO$_3$ the transverse and the longitudinal thermal conductivities peak at almost the same temperature and the thermal Hall angle respects an upper bound found in other insulators. The thermal Hall resistivity displays a weak temperature dependence over a remarkably wide temperature range. We showed that the amplitude of the thermal Hall resistivity can be accounted for by fundamental constants and the binding energy density of the solid. In contrast, the thermal Hall resistivity of dirty and complex oxides is temperature-dependent and much larger, indicative of the presence of other, yet to be identified, sources. 
	
	\section{Data availability}
	The data that support the findings of this study are available from the corresponding authors upon reasonable request.
	
	\section{Acknowledgments}
	This work is part of a Cai Yuanpei Franco-Chinese program (No. 51258NK). It was supported by the National Key Research and Development Program of China (Grant Nos. 2023YFA1609600, 2024YFA1611200 and 2022YFA1403500), the National Science Foundation of China (Grant No. 12304065, 51821005, 12004123, 51861135104 and  11574097), the Fundamental Research Funds for the Central Universities (Grant No. 2019kfyXMBZ071), and the Hubei Provincial Natural Science Foundation ‌(2025AFA072).

	\clearpage

\begin{center}{\large\bf Supplementary Materials for ``Thermal Hall resistivity and transverse entropy production in a phonon gas"}\\
\end{center}

\renewcommand{\thesection}{S\arabic{section}}
\renewcommand{\thetable}{S\arabic{table}}
\renewcommand{\thefigure}{S\arabic{figure}}
\renewcommand{\theequation}{S\arabic{equation}}

\setcounter{section}{0}
\setcounter{figure}{0}
\setcounter{table}{0}
\setcounter{equation}{0}

\section*{Note 1. Negligible contribution of mobile electrons to thermal conductivity} 

\begin{figure*}[ht!]
	\centering
	\includegraphics[width=0.7\linewidth]{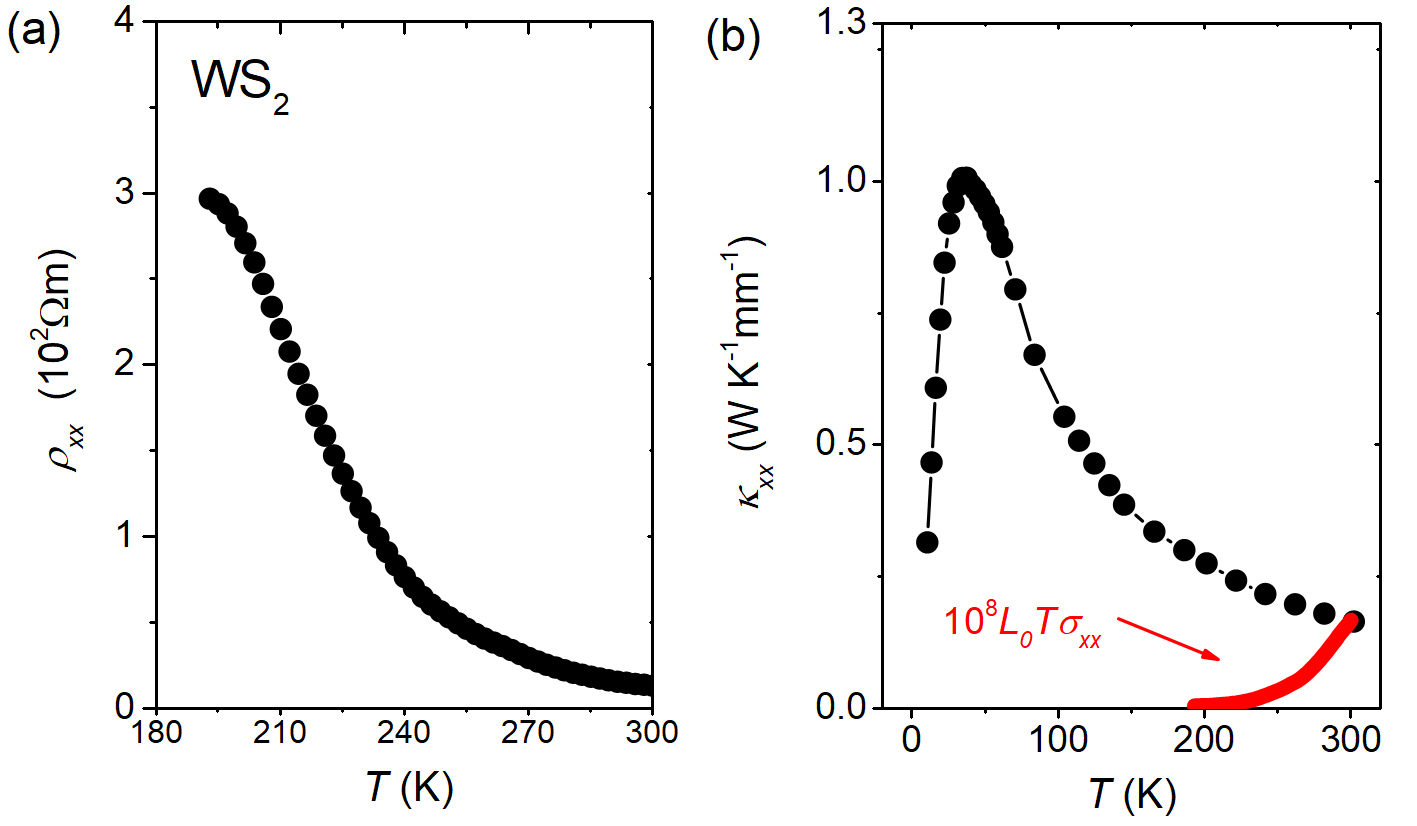} 
	\caption{\textbf{Resistivity and electronic thermal conductivity. } (\textbf{a}) The longitudinal resistivity $\rho_{xx}$ as a function of temperature $T$. (\textbf{b}) The electron thermal conductivity $\kappa_{xx}^e$, estimated using the Wiedemann–Franz law, is compared with the total thermal conductivity $\kappa_{xx}$. }
	\label{fig.s1}
\end{figure*}

Figure \ref{fig.s1}\textbf{a} shows the resistivity of WS$_2$ in the $xy$ plane. The electronic thermal conductivity $\kappa_{xx}^e$ estimated from $\rho_{xx}$ through the Wiedemann-Franz law, is about 8 orders of magnitude smaller than the total thermal conductivity $\kappa_{xx}$ as seen in Fig.\ref{fig.s1}\textbf{b}, implying that phonons dominate the thermal transport in WS$_2$ \cite{pisoni2016anisotropicSM}. Thus, it is safe to neglect  the electronic contribution to heat transport.

\section*{Note 2. Experimental setup for longitudinal and transverse thermal conductivity measurements}

\begin{figure}[ht!]
	\centering
	\includegraphics[width=0.5\linewidth]{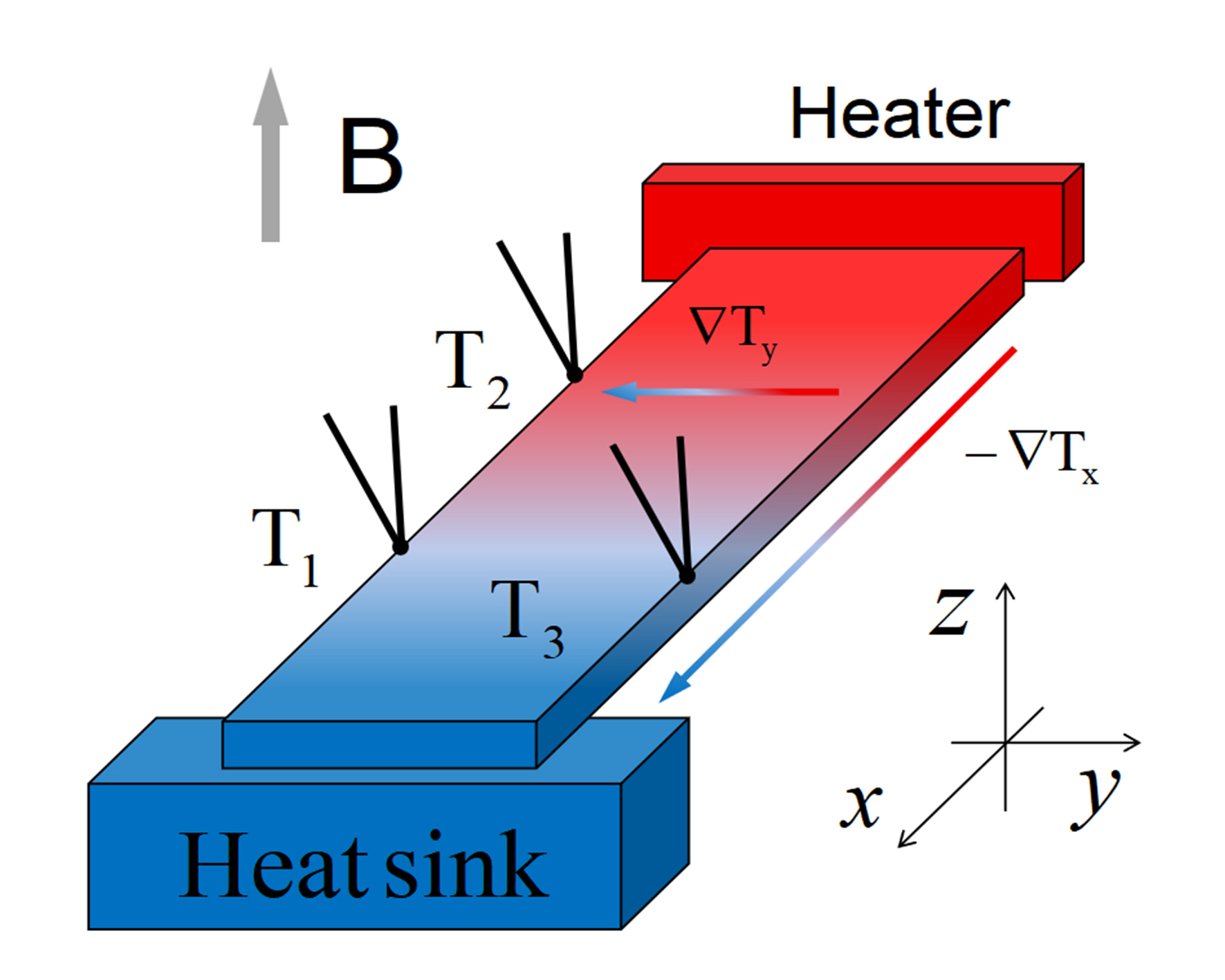} 
	\caption{\textbf{Schematic of the thermal-transport measurement configuration.}
		A longitudinal heat current is generated by the heater and flows along the $x$ direction toward the heat sink. Three thermocouples, $T_1$, $T_2$, and $T_3$, are used to determine the longitudinal and transverse temperature gradients.}
	\label{fig.s2}
\end{figure}

Figure~\ref{fig.s2} illustrates the configuration used for the longitudinal and transverse thermal-transport measurements. The longitudinal heat current  generated by the heater flows along the $x$ direction of the sample toward the heat sink, parallel to $-\nabla_x T$. Three type-E thermocouples, denoted $T_1$, $T_2$, and $T_3$, were used to determine the longitudinal and transverse temperature gradients. The magnetic field was applied along the $z$ direction, perpendicular to the sample plane, while the transverse temperature gradient $\nabla_y T$ was measured along the $y$ direction \cite{Li2023SM}.

At each stabilized magnetic field, the measurement followed a three-step sequence. We first recorded the background voltage with the heater switched off, then measured the thermocouple signal with the heater switched on, and finally recorded a second background after switching the heater off again. The  temperature gradient in the presence of the heat flow was obtained after subtracting the average background in the absence of heat flow. 

\section*{Note 3. Processing the raw data}

\begin{figure*}[ht!]
	\centering
	\includegraphics[width=0.9\linewidth]{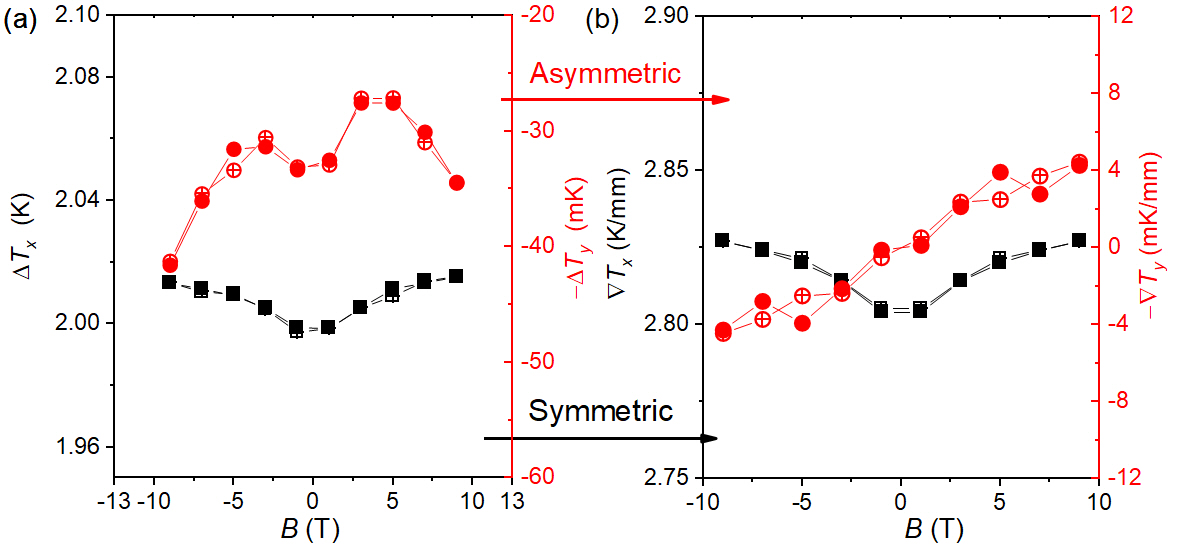} 
	\caption{\textbf{Raw data and their processing. } (\textbf{a}) The longitudinal and transverse temperature differences at 23.3 K under a heating power of 40 mW. (\textbf{b}) The results after the symmetric and asymmetric processing. }
	\label{fig.s3}
\end{figure*}

Fig. \ref{fig.s3} shows the longitudinal and transverse temperature differences at 23.3 K upon the application of a heating power of 40 mW in WS$_2$ sample \#1. The magnetic field was swept from $-9$ to $+9$ T. The longitudinal temperature difference $\Delta T_{x}$ exhibits predominantly symmetric behavior. It is almost totally even in the magnetic field. In contrast, the transverse temperature difference $\Delta T_{y}$ is predominantly antisymmetric behavior. The mostly odd signal is accompanied by a small even background. To separate the two, we performed symmetric and asymmetric processing for the longitudinal and transverse temperature differences. For symmetric processing, we used [$\Delta T_{x}(+B) + \Delta T_{x}(-B) $]/2$l$, where $l$ is the length between two longitudinal thermocouples. For asymmetric processing, we used [$\Delta T_{y}(+B) - \Delta T_{y}(-B) $]/2$w$ \cite{Li2023SM}, where $w$ is the length between two transverse thermocouples. The processed results are shown in Supplementary Fig. \ref{fig.s3}\textbf{b}. 

\begin{figure*}[ht!]
	\centering
	\includegraphics[width=1.0\linewidth]{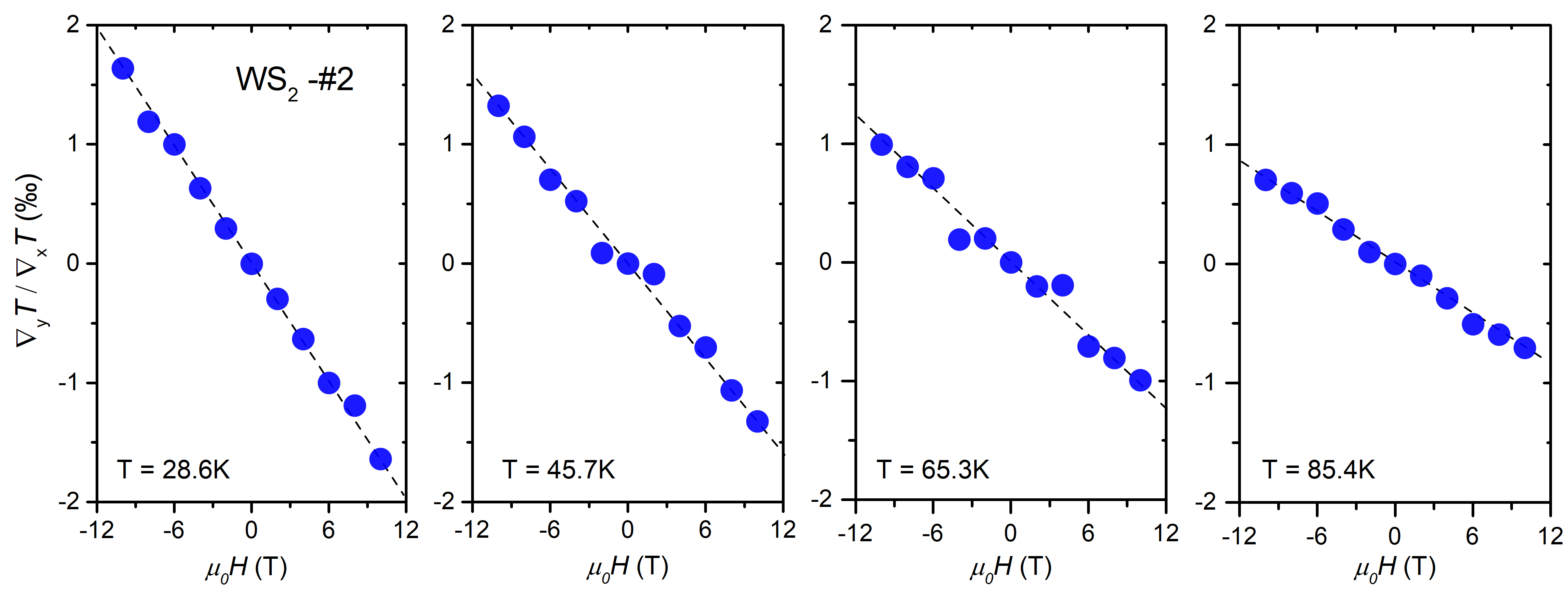} 
	\caption{\textbf{Raw temperature-dependent data for WS$_2$ sample \#2 measured at 10 T.} Thermal Hall angle $\nabla_y T / \nabla_x T $ at 28.6 K, 45.7 K, 65.3 K and 85.4 K.  }
	\label{fig.s5}
\end{figure*}

Figures \ref{fig.s4} and \ref{fig.s5} show the field dependence of the thermal Hall angle ($\nabla_y T / \nabla_x T$) for samples \#1 and \#2 at different temperatures, measured at 9 T and 10 T, respectively. For both samples, the heat current $J_q$ was applied along the $x$ axis ([11$\bar{2}$0]), and the transverse temperature gradient $(\nabla T)_\perp$ was measured along the $y$ axis ([0$\bar{1}$10]).

\begin{figure*}[ht!]
	\centering
	\includegraphics[width=1.0\linewidth]{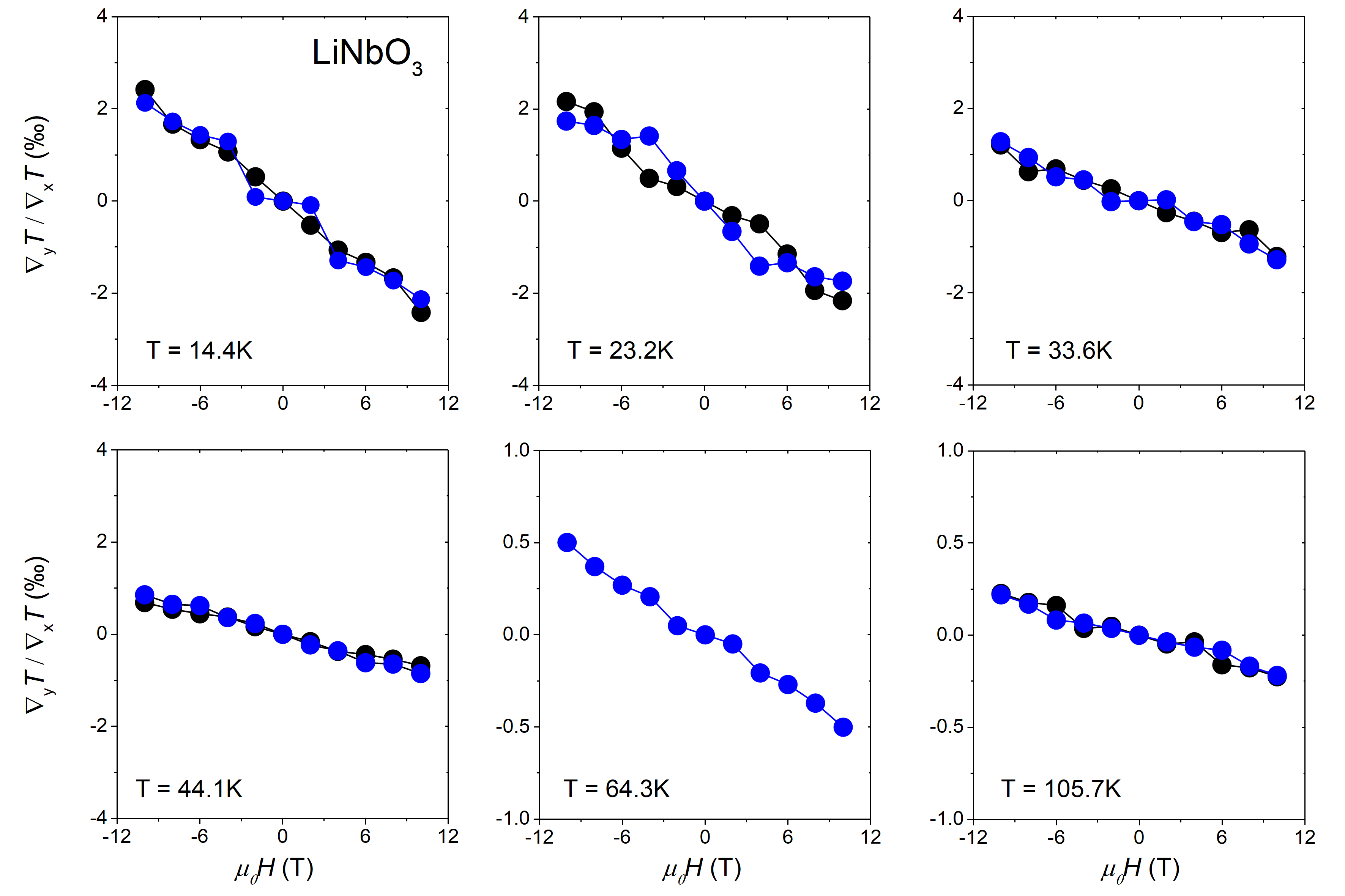} 
	\caption{\textbf{Raw temperature-dependent data for the LiNbO$_3$ sample measured at 10 T.} The thermal Hall angle $\nabla_y T / \nabla_x T$ is shown at selected temperatures of 14.4, 23.2, 33.6, 44.4, 64.3, and 105.7 K.}
	\label{fig.s6}
\end{figure*}

Following the same procedure, we measured and analyzed the thermal Hall effect of the LiNbO$_3$ single crystal. The raw data are presented in Fig.~\ref{fig.s6}, where the heat current $J_q$ is applied along the [11$\bar{2}$0] direction and the transverse temperature gradient $(\nabla T)_{\perp}$ is measured along the [0$\bar{1}$10] direction at six different temperatures.

\section*{Note 4. Specific Heat of WS$_2$ and LiNbO$_3$ Single Crystals} 

\begin{figure*}[ht!]
	\centering
	\includegraphics[width=1.0\linewidth]{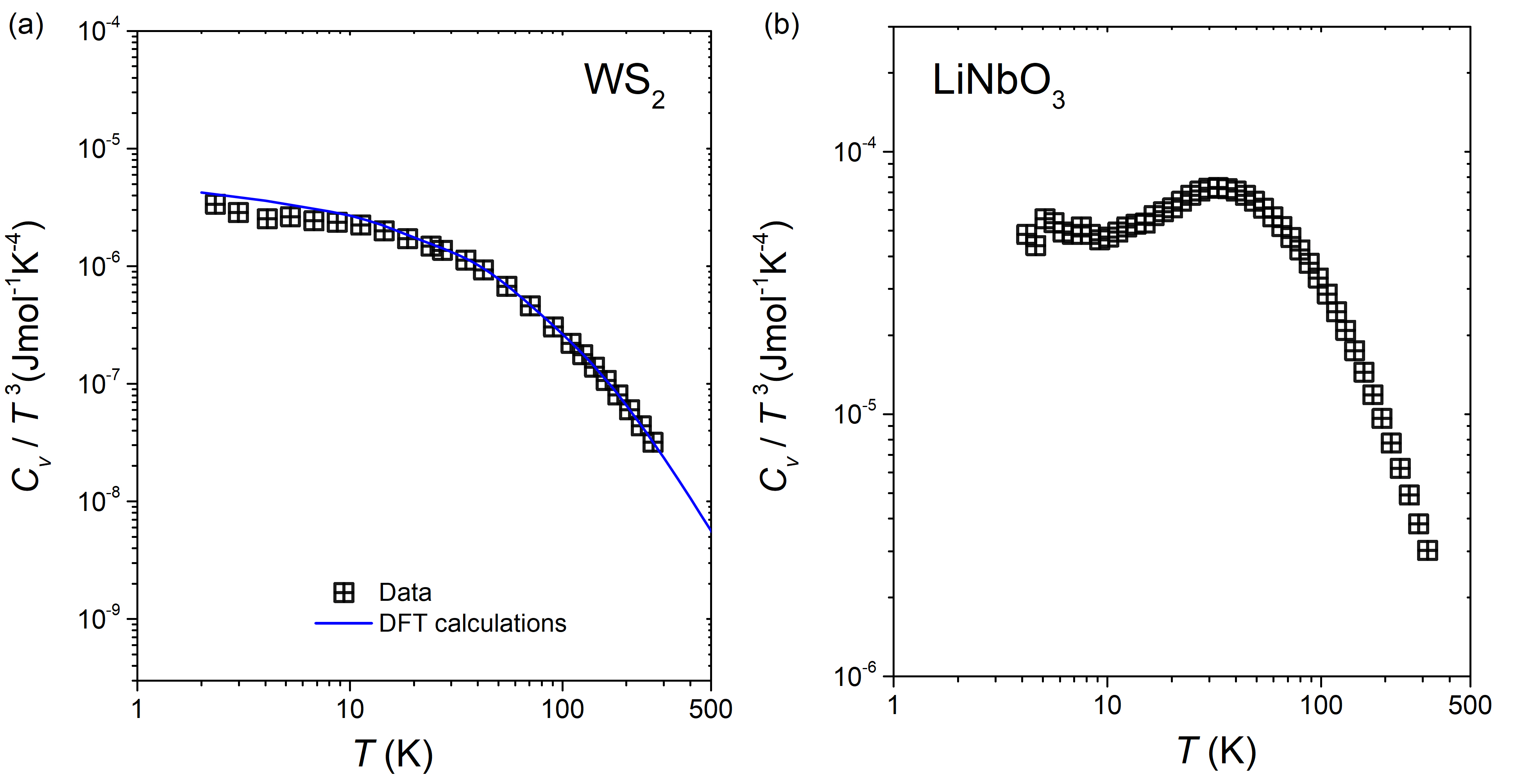} 
	\caption{\textbf{Temperature dependence of the reduced heat capacity.} $C_V/T^3$ as a function of temperature for (\textbf{a}) WS$_2$ and (\textbf{b}) LiNbO$_3$, plotted on logarithmic scales. In (\textbf{a}), the open squares represent the experimental data and the blue solid line denotes the DFT calculations. The weak temperature dependence of $C_V/T^3$ at low temperatures is consistent with the Debye relation $C_V\propto T^3$. LiNbO$_3$ exhibits a broad maximum in $C_V/T^3$ at intermediate temperatures.}
	\label{fig.s7}
\end{figure*}

Figure~\ref{fig.s7} shows the temperature dependence of heat capacity divided by the cube of temperature, $C_V/T^3$, in WS$_2$ and LiNbO$_3$. For WS$_2$, the DFT calculations reproduce both the magnitude and the temperature dependence of the experimental data over the whole measured temperature range. At low temperatures, $C_V/T^3$ approaches a constant, consistent with the Debye relation $C_V \propto T^3$ and with previous measurements \cite{specificWS2-1SM}, as expected at temperatures well below the Debye temperature.

The temperature dependence of $C_V/T^3$ in LiNbO$_3$ displays a peak at $T \approx$ 40 K \cite{PerezEnciso1998SM}, indicative of the presence of an Einstein mode on top of the Debye background. This was previously observed, for example, in elemental antimony \cite{jaoui2022SM} or strontium titanate \cite{McCalla2019SM}.

\section*{Note 5. Calculated WS$_2$ phonon spectrum} 

\begin{figure*}[ht!]
	\centering
	\includegraphics[width=0.8\linewidth]{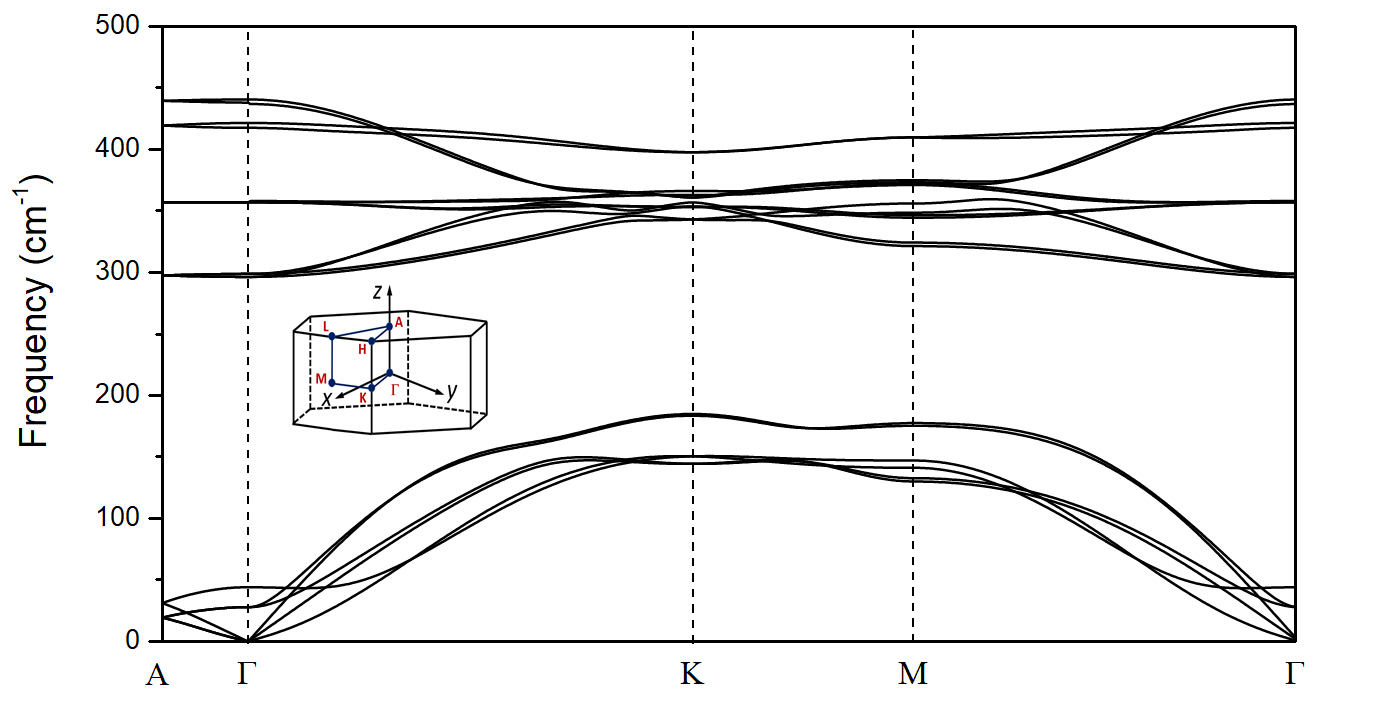} 
	\caption{\textbf{The phonon dispersions of WS$_2$. } 
		Calculated phonon dispersion of WS$_2$ along the $A$-$\Gamma$-$K$-$M$-$\Gamma$ path. The relatively weak dispersion of most branches along the out-of-plane direction $A$-$\Gamma$ reflects the layered character of the crystal. The large gap between lower and higher manifolds is one of the characteristic features of WS$_2$, which results from the large mass difference between W and S atoms.  This dispersion is in excellent agreement with what was reported in Ref.~\cite{Souliou2025SM}.}
	\label{fig.s8}
\end{figure*}

Fig. \ref{fig.s8} shows the calculated phonon dispersions of WS$_2$ along the $A$-$\Gamma$-$K$-$M$-$\Gamma$ path. The overall spectrum agrees well with the results reported by Souliou et al. and by Molina-Sanchez and Wirtz \cite{Souliou2025SM,Molina2011SM}. It contains a low-energy manifold of three acoustic and optical branches that is separated from a higher-lying manifold of twelve optical branches by a large gap. This large separation between the lower and upper parts of the phonon spectrum occurs due to the large mass difference between W and S atoms and is also consistent with the results of Lindroth and Erhart \cite{Lindroth2016SM}. The comparatively weak dispersion of many branches along the out-of-plane direction $A$-$\Gamma$ reflects the layered character of the material. The longitudinal sound velocity was estimated from the slope of the longitudinal acoustic phonon branch near the $\Gamma$ point, $v_{\mathrm{LA}}=\left.\partial\omega_{\mathrm{LA}}/\partial q\right|_{q\rightarrow0}$, to be approximately $5.7~\mathrm{km~s^{-1}}$.

This phonon dispersion was calculated within density functional perturbation theory as implemented in {\sc quantum espresso} \cite{qeSM}.  The crystal structure was first relaxed within the local-density approximation including spin-orbit coupling, using fully relativistic {\sc pslibrary} pseudopotentials \cite{pslibrarySM}, a plane-wave cutoff of 90 Ry for the wave functions and 900 Ry for the charge density, and a $12 \times 12 \times 4$ Brillouin-zone sampling grid. The relaxed lattice parameters are $a$ = 3.130 and $c$ = 12.151 \AA. Harmonic dynamical matrices were then computed on a $12 \times 12 \times 4$ $q$-point mesh.  The phonon dispersion was obtained using Fourier interpolation.  The lattice heat capacity was evaluated within the harmonic approximation from the calculated phonon spectrum by direct summation over phonon branches on a dense $48 \times 48 \times 16$ $q$-point mesh.

\section*{Note 6. Righi-Leduc effect in a free electron gas }

The Hall resistivity of the free electron gas is given by \cite{sondheimer1948theorySM}: 

\begin{equation}
	\rho_{xy}
	\equiv\frac{E_y}{J_x}= \frac{B}{ne}.
	\label{Hall-e}
\end{equation}

This celebrated expression, which links the magnitude of Hall resistivity to the inverse of electron density $n$, is rigorously derived by solving the Boltzmann equation and inverting the matrix of electric conductivity \cite{sondheimer1948theorySM}. It can also be obtained by a shortcut. The charge current flux along $x$ is $J_x=nev_d$, where $v_d$ is the drift velocity. The Lorentz force, $F_L=eBv_d$, along $y$ is countered by a Coulomb force $F_c=eE_y$. Setting $F_C=F_L$, the combination of these three expressions yields Equation \ref{Hall-e}.

The same electron gas displays a Righi-Leduc effect, the thermal counterpart of the electric Hall effect. According to the Wiedemann-Franz law,  the thermal Hall resistivity is linked to the electrical Hall resistivity, $\rho_{xy}=\frac{\pi^2}{3}\frac{k_B^2}{e^2}T W_{xy}$ \cite{chester1961lawSM}. Therefore:

\begin{equation}
	W_{xy} 
	\equiv\frac{\nabla_y T}{J^q_x}= \frac{3}{\pi^2}\frac{1}{nk_BT}\frac{eB}{k_B} 
	\label{Righi-Leduc}
\end{equation}

According to both Equation 1 of the main text and Equation \ref{Righi-Leduc}, the field-normalized thermal Hall resistivity is given by $e/k_B$ divided by an energy density. 

\section*{Note 7. Phonon pseudomomentum balance in a finite magnetic field.}

The Boltzmann equation for a phonon gas contains Normal and Umklapp collision integrals on its right hand side: 

\begin{equation}
	\frac{\partial n_{\mathbf q\lambda}}{\partial t}
	+
	\mathbf v_{\mathbf q\lambda}\cdot\nabla T\,
	\frac{\partial n^0_{\mathbf q\lambda}}{\partial T}
	=
	C_N[n_{\mathbf q\lambda}]
	+
	C_U[n_{\mathbf q\lambda}] ,
\end{equation}

In the Callaway \cite{Callaway1959SM} description, the displaced Bose distribution is a zero mode of the Normal collision operator because a Normal three-phonon process conserves crystal momentum. 

In the presence of a finite magnetic field, however, a three-phonon event involves not only the change of phonon crystal momentum but also the change of the electromagnetic pseudomomentum associated with the lattice displacement and its induced polarization. Following the Minkowski description of electromagnetic momentum in matter \cite{Nelson1991SM}, the total pseudomomentum of the phonon--electromagnetic--lattice system should therefore be considered in an elementary process \(1+2\leftrightarrow3\). The corresponding collision integral may be written schematically as
\[
C_N[n;B]=C_N^{(0)}[n]+\delta C_N[n;B],
\]
\[
\qquad
\delta C_N[n;B]\propto
\Delta {\bf P}_{\rm M}^{\,123}(B),
\]

Here \(\Delta {\bf P}_{\rm M}^{\,123}\) denotes the change of the electromagnetic/material pseudomomentum associated with the displacement fields before and after the three-phonon event. In a Normal process,  \({\bf q}_1+{\bf q}_2={\bf q}_3\), but the corresponding pseudomomentum balance depends on the magnetic field.

A microscopic treatment would therefore require the three-phonon transition probability, and hence the collision integral on the right-hand side of the Boltzmann equation, to be evaluated with the magnetic-field-dependent electromechanical states of the initial and final phonons. In particular, one expects a transition kernel, odd in magnetic field:
\begin{equation}
	W_{123}(B)=W_{123}^{(0)}+B\,W_{123}^{(1)}+\mathcal O(B^2),
\end{equation}

It generates a transverse pseudomomentum transfer. Such a calculation could reveal how nominally momentum-conserving Normal processes acquire a transverse component in a magnetic field and thereby generate a thermal Hall response. Developing this field-dependent three-phonon collision kernel from first-principles electronic structure and lattice dynamics emerges as a challenge for future \emph{ab initio} calculations of the phonon thermal Hall conductivity in specific materials.

\begin{figure*}[ht!]
	\centering
	\includegraphics[width=0.9\linewidth]{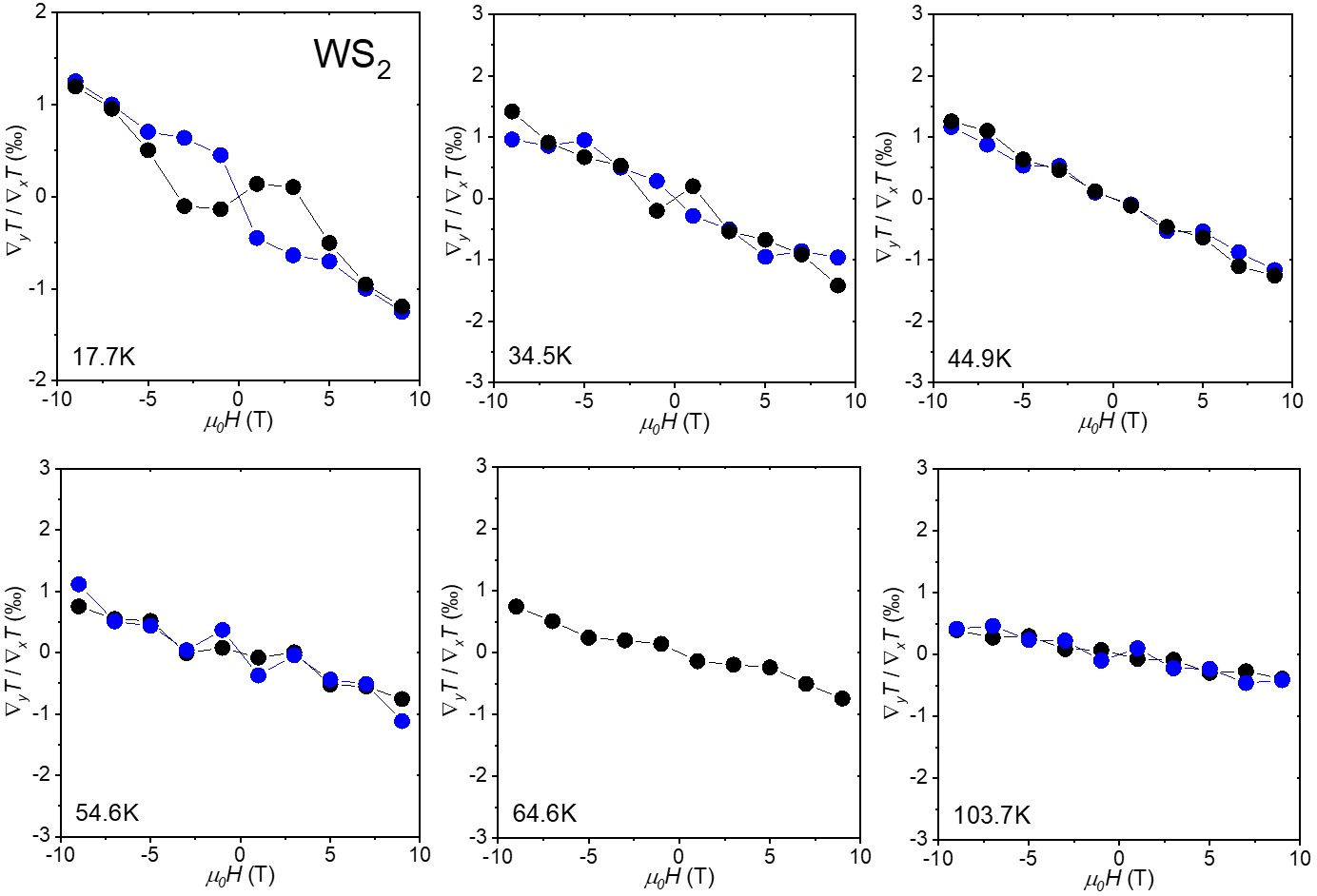} 
	\caption{\textbf{Raw temperature-dependent data for WS$_2$ sample \#1 measured at 9 T.} Thermal Hall angle $\nabla_y T / \nabla_x T $ at 17.7 K, 34.5 K, 44.9 K, 54.6 K, 64.6 K, and 103.7 K.  }
	\label{fig.s4}
\end{figure*}

\clearpage
%

\end{document}